\documentclass[a4paper,12pt]{article}
\usepackage{graphicx}
\begin{document}
\newcommand{\mz}{m_Z}
\newcommand{\mhii}{m_{H_2}}
\newcommand{\mer}{m_{{\tilde{e}_R}}}
\newcommand{\mel}{m_{{\tilde{e}_L}}}
\newcommand{\msnu}{m_{{\tilde{\nu}}}}
\newcommand{\msq}{m_{{\tilde{q}}}}
\newcommand{\ser}{\tilde{e}_R}
\newcommand{\sel}{\tilde{e}_L}
\newcommand{\snu}{\tilde{\nu}_L}
\newcommand{\snue}{\tilde{\nu}_e}
\newcommand{\snut}{\tilde{\nu}_{\tau}}
\newcommand{\sql}{\tilde{q}_L}
\newcommand{\sul}{\tilde{u}_L}
\newcommand{\sdl}{\tilde{d}_L}
\newcommand{\stau}{\tilde{\tau}}
\newcommand{\staul}{\tilde{\tau}_1}
\newcommand{\stauh}{\tilde{\tau}_2}
\newcommand{\mgl}{m_{{\tilde{g}}}}
\newcommand{\mupl}{m_{{\tilde{u}_L}}}
\newcommand{\mupr}{m_{{\tilde{u}_R}}}
\newcommand{\msql}{m_{{\tilde{q}_L}}}
\newcommand{\mdnl}{m_{{\tilde{d}_L}}}
\newcommand{\mdnr}{m_{{\tilde{d}_R}}}
\newcommand{\mlsp}{m_{{\tilde{\chi}^0_1}}}
\newcommand{\mzii}{m_{{\tilde{\chi}^0_2}}}
\newcommand{\mziii}{m_{{\tilde{\chi}^0_3}}}
\newcommand{\mziv}{m_{{\tilde{\chi}^0_4}}}
\newcommand{\mtaul}{m_{\tilde{\tau}_1}}
\newcommand{\mtauh}{m_{\tilde{\tau}_2}}
\newcommand{\mntau}{m_{\tilde{\nu}_{\tau}}}
\newcommand{\mbl}{m_{\tilde{b}_1}}
\newcommand{\mbh}{m_{\tilde{b}_2}}
\newcommand{\mtl}{m_{\tilde{t}_1}}
\newcommand{\mth}{m_{\tilde{t}_2}}
\newcommand{\ziv}{\tilde{\chi}^0_4}
\newcommand{\ziii}{\tilde{\chi}^0_3}
\newcommand{\wii}{\tilde{\chi}^+_2}
\newcommand{\wi}{\tilde{\chi}^+_1}
\newcommand{\mwi}{m_{\tilde{\chi}^+_1}}
\newcommand{\mwii}{m_{\tilde{\chi}^+_2}}
\newcommand{\lsp}{\tilde{\chi}^0_1}
\newcommand{\zii}{\tilde{\chi}^0_2}
\newcommand{\sq}{\tilde{q}}
\newcommand{\tchi}{\tilde{\chi}}
\newcommand{\gsim}{\buildrel>\over{_\sim}}
\newcommand{\lsim}{\buildrel<\over{_\sim}}
\newcommand{\psla}{p\kern-.45em/}
\newcommand{\esla}{E\kern-.45em/}
\newcommand{\tl}{\tilde{l}}

\setcounter{page}{0}
\thispagestyle{empty}
\begin{flushright}
UT-ICEPP 00-05 \\
YITP--00--38 \\
TUM--HEP--380--00 \\
July 2000 \\
\end{flushright}

\vspace{2cm}

\begin{center}
{\Large \bf Scrutinizing LSP Dark Matter at the LHC}

\baselineskip=32pt

Manuel Drees$^a$, Yeong Gyun Kim$^b$, Mihoko M. Nojiri$^b$, \\ Daisuke Toya$^c$, 
Kazumi Hasuko$^c$ and Tomio Kobayashi$^c$

\baselineskip=22pt

{\it $^a$\, Physik Department, TU M\"unchen, D--85748 Garching, Germany} \\
{\it $^b$\, YITP, Kyoto University, Kyoto, 606-8502, Japan} \\
{\it $^c$\, ICEPP, University of Tokyo, Hongo, Bunkyo, Tokyo, 113-0033, Japan}

\end{center}

\vspace{1cm}

\begin{abstract}

We show that LHC experiments might well be able to determine all the
parameters required for a prediction of the present density of thermal
LSP relics from the Big Bang era. If the LSP is an almost pure bino we
usually only need to determine its mass and the mass of the $SU(2)$
singlet sleptons. This information can be obtained by reconstructing
the cascade $\tilde{q}_L \rightarrow \tilde{\chi}_2^0 q \rightarrow
\tilde{\ell}_R \ell q \rightarrow \tilde{\chi}_1^0 \ell^+ \ell^-
q$. The only requirement is that $m_{\tilde{\ell}_R} <
m_{\tilde{\chi}_2^0}$, which is true for most of the cosmologically
interesting parameter space. If the LSP has a significant higgsino
component, its predicted thermal relic density is smaller than for an
equal--mass bino. We show that in this case squark decays also
produce significant numbers of $\tilde{\chi}_4^0$ and
$\tilde{\chi}_2^\pm$. Reconstructing the corresponding decay cascades
then allows to determine the higgsino component of the LSP.

\end{abstract}

\vspace{2cm}

\vfill

\pagebreak

\baselineskip=14pt

\section{Introduction}

The Minimal Supersymmetric Standard Model (MSSM) \cite{SUSY} is one of
the most promising extensions of the Standard Model. It offers a
natural solution of the hierarchy problem \cite{witten}, amazing gauge
coupling unification \cite{amaldi}, and Dark Matter candidates
\cite{jungman}. If Nature chooses low energy supersymmetry (SUSY),
sparticles will be found {\em for sure}, as they will be copiously
produced at future colliders such as the Large Hadron Collider (LHC)
at CERN or TeV scale $e^+e^-$ linear colliders (LC). The LHC would be
a great discovery machine if SUSY breaking masses lie below a
few TeV \cite{snowmass}. On the other hand, there are several on-going
and future projects searching for LSP Dark Matter. One of them even
claims a positive signal \cite{DAMA}, although the current situation
is rather contradictory \cite{CDMS}. In any case, it seems very
plausible that both SUSY collider signals and LSP Dark Matter in the
Universe will be found in future.

Recently interesting possibilities have been pointed out where
non--thermal production of Dark Matter is significant \cite{RM,EM}.
Generally the known bound from the thermal LSP density may easily be
evaded by assuming a low post--inflationary reheating temperature of
the Universe, without endangering the standard successes of Big Bang
cosmology \cite{G}. If the reheating temperature is below the
neutralino decoupling temperature, the relation between neutralino
pair annihilation rates and the mass density of the Universe
disappears. The mass may then be determined by other parameters, such
as the Q ball formation rate and decay time \cite{EM}, or the moduli
masses and their decay rates to LSPs \cite{RM}.

While these non--thermal mechanisms open exciting new possibilities,
direct experimental or observational tests of them might be difficult,
since they all have to occur before Big Bang nucleosynthesis
(BBN).\footnote{For the Q ball case, measurements of the cosmic
microwave background by the MAP and PLANCK satellites might find
isocurvature fluctuations due to the Affleck--Dine Field \cite{EM1}.}
It is therefore interesting to determine
\begin{enumerate}
\item  the actual LSP relic density, both ``locally'' (in the solar
system) and averaged over the Universe; and
\item  the predicted thermal LSP relic density,
\end{enumerate}
as precisely as possible. These quantities are closely related to the
mass and interactions of the LSP. A positive difference between the
actual and predicted LSP density would indicate the existence of
non--thermal relics, whereas a negative difference would hint at large
entropy production below the LSP freeze--out temperature (e.g. due to
a low reheating temperature).

The matter density in the Universe divided by the critical density,
$\Omega_{\rm matter}$, is claimed to be tightly constrained already;
$\Omega_{\rm matter} = 0.35 \pm 0.07 $ \cite{MT}. On the other hand, the
thermal relic density of the Universe $\Omega_{\lsp} h^2$
($h=0.65\pm0.05$) has been calculated through the mass and interaction
of the LSP, which is likely to be the lightest neutralino $\lsp$. In
the absence of direct production of sparticles we have to rely on
experimental lower bounds on sparticle masses as well as naturalness
arguments to conclude that the predicted $\Omega_{\lsp} h^2$ lies
somewhere between $10^{-3}$ and $10^3$; clearly this is not a very
useful prediction, although it is encouraging that this wide range at
least includes the correct value.  The purpose of this paper is to
discuss how future LHC experiments can contribute to the determination
of the MSSM parameters that are needed to predict the thermal LSP
relic density and the LSP--nucleon scattering cross section. Our goal
is thus somewhat different from that of ref.\cite{kane}, where it was
simply assumed that all relevant parameters had somehow been
determined by various experiments, with given errors; the main
emphasis was on estimating the resulting uncertainties in the
predictions of the thermal LSP relic density and the LSP--nucleon
scattering cross section. In contrast, we discuss in some detail {\em
how} these parameters can be determined, and with what errors.

The determination of mass parameters has been discussed in detail in
the minimal supergravity (mSUGRA) model, where one assumes
universality of scalar masses and of gaugino masses at the scale $M_X$
of Grand Unification \cite{HP}. In Sec.~2, we point out that
$\zii\rightarrow \ser$ is open for most parameters giving a reasonable
LSP density, making the determination of $\mzii$, $\mlsp$ and $\mer$
possible at the LHC. We demonstrate that the mass density is
determined by the LSP and slepton masses, if the LSP is mostly a bino
as expected in mSUGRA. In this case $\Omega_{\lsp}$ can be
predicted to about 10 to 20\% accuracy.

In Secs.~3 and 4 we discuss a non--mSUGRA scenario. In Sec.~3 we relax
the assumption of universal scalar masses for Higgs bosons. It is then
easy to find cases with comparable higgsino and gaugino masses,
$\mu\sim M$, while keeping all squared scalar masses positive at
$M_X$. The LSP then has a significant higgsino component, so that its
density cannot be predicted by only studying $\zii \rightarrow \ser
\rightarrow \lsp$ decays. The situation is further complicated if we
also relax the assumption of universal gaugino masses, since the
neutralino mass matrix then depends on three independent, unknown mass
parameters. We point out that the cascade decay $\wii \rightarrow \snu
\rightarrow \wi$ can then often be identified, providing clear
evidence that $\mu\sim M$. In Sec.~4 we present a detailed case study
with $\mu\sim M_2$ to confirm the potential of LHC experiments to
analyze $\wii$ cascade decays; this allows a complete determination of
the neutralino mass matrix (in the absence of CP--violating
phases). Sec.~5 is devoted to discussions.

\section{$\Omega_{\lsp}$ in mSUGRA}

In the minimal supergravity model one assumes universal soft breaking
parameters at the GUT scale: a universal scalar mass $m$, universal
gaugino mass $M$, universal trilinear coupling $A$, and Higgs mass
parameter $B$. The renormalization group evolution of soft breaking
squared Higgs masses then leads to consistent breaking of the
electroweak symmetry, provided the higgsino mass parameter $\mu$ can
be tuned independently. In this paper, we chose the weak scale input
parameters $m_b(m_b)=4.2$ GeV, $m_t(m_t)=165$ GeV, and $\tan\beta$. We
minimize the tree level potential at renormalization scale $Q=
\sqrt{m_{\tilde{t}}m_t}$, which reproduces the correct value of $\mu$
obtained by minimizing the full 1--loop effective potential. We
include loop corrections to the masses of neutral Higgs bosons,
including leading two--loop corrections \cite{2loop}. 

The mass density $\Omega_{\lsp} h^2$ of the LSP is calculated from the
pair annihilation cross section by using the expressions \cite{GKT}
\begin{eqnarray}
\Omega_{\lsp} h^2&=& \frac{1.07\times 10^9/{\rm GeV} x_F }
{\sqrt{g_*}M_P(a+3b/x_F)} \cr
\sigma(\lsp\lsp\rightarrow {\rm all})v &=& a+bv^2\cr
x_F(\equiv \mlsp/T_F) &=&\log 
\frac{\left(0.764 M_P(a+6b/x_F) c(2+c)\mlsp\right)}{\sqrt{g_* x_F}}
\end{eqnarray}
except near regions of parameter space where special care is needed.
Here $M_P=1.22\cdot 10^{19}$ GeV is the Planck mass, and $a$ and $b$
are the first two coefficients in the Taylor expansion of the pair
annihilation cross section of the LSP with respect to the relative
velocity $v$ of the LSP pair in its center of mass frame. $g^*$ is the
effective number of relativistic degrees of freedom at LSP freeze--out
temperature $T_F$. The expansion in $v$ breaks down around $s-$channel
poles; here the thermal average is calculated numerically, using the
formalism given by Griest and Seckel \cite{GS}. We also take into
account sub--threshold annihilation into $hh$ and $W^+W^-$ final
states.  When the LSP is higgsino--like, coannihilation of $\lsp$ with
$\wi$ and $\zii$ are important \cite{MY,DN3}. The annihilation modes
$\lsp\wi \rightarrow ff'$ and $\lsp \zii \rightarrow f \bar{f}$ are
approximated by $s-$channel $W$ and $Z$ exchange, respectively;
$\lsp\wi\rightarrow W\gamma$ is also included. All other higgsino
coannihilation modes are treated assuming $SU(2)$ invariance.  We also
include one loop radiative corrections to the mass splitting of
higgsino--like states \cite{dp}. We do not include $\lsp\tilde{\tau}$
coannihilation, since we do not study cases with $m_{\lsp} \simeq
m_{\tilde{\tau}}$.

MSUGRA predicts a bino--like LSP $\lsp$ and wino--like $\wi$ and
$\zii$ for moderate values of $m$ and $M$ (below $\sim 500$ GeV). This
is a rather model independent result \cite{falk}. Large positive
corrections to squark masses from gaugino loops, together with the
large top Yukawa coupling, drive the squared soft breaking Higgs mass
$m^2_2$ negative at the weak scale. On the other hand, correct
symmetry breaking requires $m^2_2+\mu^2> -\mz^2/2$. One has to make
$\mu$ large to obtain the correct electroweak symmetry breaking scale,
if scalar masses and gaugino masses are of the same order.

If slepton masses are moderate, the LSP is bino--like, and one is
sufficiently far away from $s-$channel poles ($2 m_{\lsp} \neq m_Z,
m_{\rm Higgs}$), the mass density is essentially determined by
$t-$channel $\ser$ exchange \cite{DN3}. This is because
\begin{enumerate}
\item  A pure bino couples only to fermion and sfermion, or Higgs 
and higgsino. Higgsino exchange is suppressed for $\mu^2\gg M^2$.  
\item  $\mer < \mel \simeq \msnu \ll \msq$ in mSUGRA, therefore $\ser$ 
exchange is least suppressed by large masses in the propagator.
\item The hypercharges of sleptons satisfy the relation
$Y_{\ser}=2Y_{\sel}$, therefore $\sigma(\ser)v = 16 \sigma(\sel \ {\rm
or}\ \tilde{\nu}_L)v$ when sfermion masses are equal.
\end{enumerate}
\begin{figure}[htbp]
\begin{center}
\hskip 1cm 
\includegraphics[width=6.1cm,angle=-90]{omega1.epsf}
\hskip 0cm 
\includegraphics[width=6.1cm,angle=-90]{omega2.epsf}
\end{center}
\vskip-0.3cm 
\hskip 3.5cm 
a)  
\hskip 8cm 
b)
\begin{center}
\hskip 1cm 
\includegraphics[width=6.1cm,angle=-90]{bfactor1.epsf}
\hskip 0cm 
\includegraphics[width=6.1cm,angle=-90]{bfactor2.epsf}
\end{center}
\vskip-0.3cm 
\hskip 3.5cm 
c)  
\hskip 8cm 
d)
\caption{\footnotesize Contours of constant $\Omega_{\lsp} h^2$ (Fig 1
a,b) and $b$ factor (Fig 1 c,d) in the ($m$,$M$) plane for
$\tan\beta=10$ (a,c) and 4 (b,d). We take $\mu>0$. Contours of
constant $\mlsp$, as well as contours where $\mlsp = \mer$ and $\mlsp
= \mel$ are also shown.}
\label{fig1}
\end{figure}

This can be seen in Fig.~1 a)-d), where $\Omega_{\lsp}$ (a, b) and $b
\equiv 10^6 \ {\rm GeV}^2 \times \Omega_{\lsp} h^2\sigma_{\tilde B}$ (c,d)
are plotted. Here $\sigma_{\tilde B}$ is the scaled bino pair
annihilation cross section in the limit where $\mer\ll\mel,\msq$
\cite{DN3},
\begin{equation} \label{ebino}
      \sigma_{\tilde B}=\frac{\mlsp^2}{(\mer^2+\mlsp^2)^2} \times
\left[ \left(1 - \frac{\mlsp^2}{\mer^2+\mlsp^2} \right)^2
      + \frac{\mlsp^4}{(\mer^2 + \mlsp^2)^2}\right].
\end{equation} 
We find that the mass density increases with increasing
$M\propto\mlsp$ and $m$; $m$ is essentially proportional to $\mer$ for
$m\geq M$. Dotted lines are for constant $\lsp$ mass. The fact that it
basically only depends on $M$ indicates that the LSP is indeed
bino--like. The mass density becomes very small for $\mlsp=m_Z/2$,
because LSP pair annihilation through $Z$ exchange is enhanced. In
Fig.~1 c) and 1d) we show contours of constant $b$. Although
$\Omega_{\lsp} h^2$ changes by more than a factor of 4, the change of $b$ is
very small over the wide range of parameter region with $M>160$ GeV,
again confirming the bino--like nature of the LSP for the mSUGRA case.

\begin{figure}[htbp]
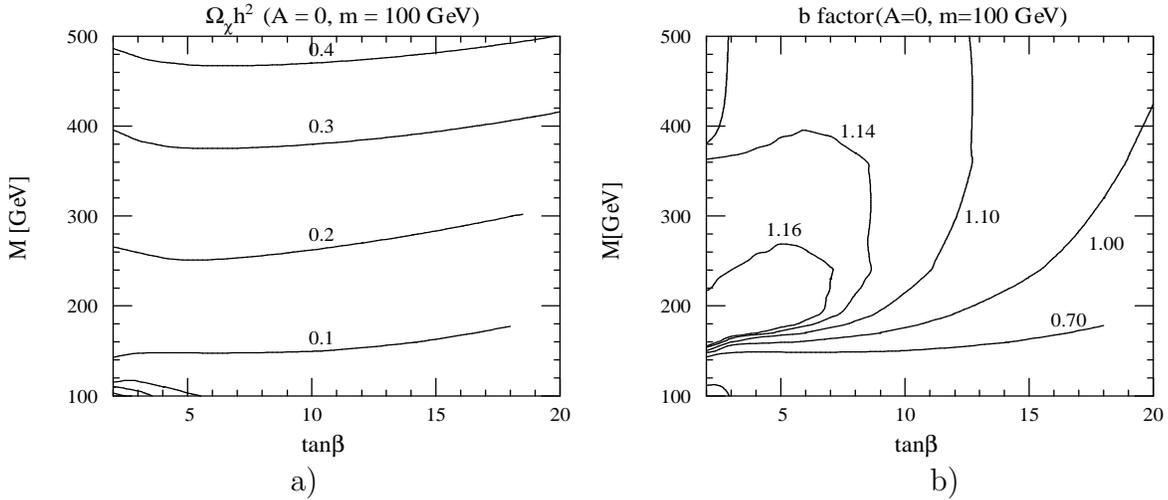

\begin{center}
\hskip 1cm 
\includegraphics[width=6.1cm,angle=-90]{omega.epsf}
\hskip 0cm 
\includegraphics[width=6.1cm,angle=-90]{bfac.epsf}
\end{center}
\vskip-0.3cm 
\hskip 3.5cm 
a)  
\hskip 8cm 
b)
\caption{\footnotesize $\Omega_{\lsp} h^2$ and the $b$ factor 
in the ($\tan\beta$,$M$) plane for fixed $m$.}
\end{figure}
The $\tan\beta$ dependence is also very weak, as can be seen in Fig.~2
a) and b). This again shows that the LSP is bino dominant, and
bino--higgsino mixing, which is controlled by the off--diagonal
elements of the neutralino mass matrix, has negligible effect on the
LSP relic density for the parameters given in the plot. For
sufficiently heavy $\lsp$, $\Omega_{\lsp}$ is simply determined by $\mer$ and
$\mlsp$ so that $b\sim 1 $.

We now show that analyses of sparticle production at the LHC would
lead to tight constraints on the predicted thermal relic density
$\Omega_{\lsp} h^2$. Recently, quite a few studies of precision
measurements of sparticle masses at the LHC have been performed. When
the cascade decay $\tilde{q} \rightarrow \zii \rightarrow \ser
\rightarrow \lsp$ is open, a clean SUSY signal is $ll+ jets +$ missing
$E_T$. It was shown \cite{HP} that $\msq$, $\mzii$, $\mer$ and $\mlsp$
can be reconstructed from the upper end points of the $m_{jll}$ and
$m_{jl}$ distributions, $m^{\rm max}_{jll}$ and $m^{\rm max}_{jl}$;
the edge of the $m_{ll}$ distribution, $m^{\rm max}_{ll}$; and the
lower end point of the $m_{jll}$ distribution with $m_{ll}>m^{\rm
max}_{ll}/\sqrt{2}$, $m^{\rm min}_{jll}$. Here $j$ refers to one of
the two hardest jets in the event. In most cases it is chosen such
that it has the smaller $jll$ invariant mass; this is meant to select
the jet from the primary $\tilde{q} \rightarrow \zii q$
decay. However, $m^{\rm min}_{jll}$ is reconstructed by taking the jet
which gives the larger $jll$ invariant mass, in order to avoid
contamination. Those end points are given by the analytical formulae
\cite{HP}:
\begin{eqnarray} \label{ekin1}
m^{\rm max}_{jll} & = & \left[\frac
{(\msql^2-\mzii^2)(\mzii^2-\mlsp^2)}
{\mzii^2}
\right]^{1/2}\cr
m^{\rm max}_{jl}  & = & {\rm Max}\left(
\left[
\frac{(\msql^2-\mzii^2)(\mzii^2-\mer^2)}
{\mzii^2}
\right]^{1/2},
\ \left[
\frac{(\msql^2-\mzii^2)(\mer^2-\mlsp^2}
{\mer^2}
\right]^{1/2}
\right) 
\cr
m^{\rm max}_{ll}  & = & \sqrt{\frac
{(\mzii^2-\mer^2)(\mer^2-\mlsp^2)}{\mer^2}
}\cr
m^{\rm min}_{jll} & = & \frac{1}{4 \mzii^2 \mer^2}
\left[
-\mlsp^2 \mzii^4 +3\mlsp^2\mzii^2\mer^2-\mzii^4\mer^2 -\mzii^2\mer^4
-\mlsp^2\mzii^2\msql^2\right.\cr 
&& \left.-\mlsp^2\mer^2\msql^2 + 3 \mzii^2\mer^2\msql^2 
-\mer^4\msql^2 +(\mzii^2-\msql^2)\times 
\right.\cr
& &\left.\sqrt{
(\mlsp^4+\mer^4)(\mzii^2+\mer^2)^2 + 2 \mlsp^2\mer^2(\mzii^4-6
\mzii^2\mer^2+\mer^4)
}
\right]
\end{eqnarray}

In addition to those quantities, one can measure the end point
$m^{\rm min}_{jl}$ of the distribution of the smaller of the two $m_{jl}$
values. It can be expressed as
\begin{eqnarray} \label{ekin2}
m^{\rm min}_{jl} & =& \sqrt{\frac{(\msql^2-\mzii^2)(\mer^2-\mlsp^2)}
{2\mer^2-\mlsp^2}} \ \ {\rm if }\ \   2\mer^2-(\mlsp^2 +\mzii^2)<0
\cr
& = & \sqrt{\frac{(\msql^2-\mzii^2)(\mzii^2-\mer^2)}{\mzii^2}}
\ \ {\rm if }\ \   2\mer^2-(\mlsp^2 +\mzii^2)>0
\end{eqnarray}
Because there are only four masses involved, the last end point is
redundant, but might be useful to cross check the decay
kinematics.\footnote{We are assuming that squarks are basically
degenerate. Note that essentially only left--handed squarks will
contribute here, since $SU(2)$ singlet squarks very rarely decay into
$\zii$. Barring ``accidental'' cancellations, bounds on flavor
changing neutral current processes imply that squarks with equal gauge
quantum numbers must be close in mass. The mass splitting between
$\tilde{u}_L$ and $\tilde{d}_L$ squarks is limited by $SU(2)$
invariance. With the possible exception of third generation squarks
the assumed degeneracy therefore holds almost model--independently. If
required, contributions from third generation squarks can be filtered
out by anti--tagging $b-$jets.}

For the example studied in \cite{HP}, the so--called ``point 5'',
$m=100$ GeV and $M= 300$ GeV, which results in $\mzii=233$ GeV,
$\mer=157.2$ GeV and $\mlsp=121.5$ GeV. The errors on $\mer$ and
$\mlsp$ are strongly correlated and are found to be 12\% for $\mlsp$
and 9\% for $\mer$. Within the framework of mSUGRA the measured LSP
mass excludes the possibility that $s-$channel poles are important for
the LSP pair annihilation cross section (see below). We find that the
corresponding error on $\sigma_{\tilde{B}}$, and hence on the
prediction for $\Omega_{\lsp} h^2$, for this parameter point is 20\%. If
the error (which is dominated by systematics associated with
uncertainties of signal distributions) is reduced by a detailed study of
various signal distributions, the error on $\sigma_{\tilde{B}}$ may go
down below 10\%.

Fig.~1 also contains contours where $\mzii=\mer$ and $\mzii=\mel$. To
the left of these contours $\zii$ decays into $\tilde{l}$ are
accessible, giving a substantial constraint to the kinematics of the
events. First, note that $\zii \rightarrow \ser$ decays are open for
most of the cosmologically acceptable region with $M \geq 200$
GeV. There is even a substantial region of parameter space where both
$\zii\rightarrow\sel$ and $\zii\rightarrow\ser$ are open. In our
argument above, we assumed that all observed edges and end points
kinematic distributions come from $\sql\rightarrow \zii\rightarrow
\ser$ rather than $\zii\rightarrow\sel$. When the latter decay mode is
open, the branching ratio dominates over $\ser$, since $\zii \simeq
\widetilde{W}_3$ and $\ser$ is an $SU(2)$ singlet. However if one
assumes that squared scalar masses are positive at the GUT scale, $\sel$
cannot be too much lighter than $\zii$, so there is some kinematical
suppression. In mSUGRA the relevant masses are expressed as:
\begin{eqnarray}
M_1&=& 0.4 M,\ \ \ \  M_2=0.8 M\cr
\mel^2 &=& m^2+ 0.5 M^2 - \left(\frac{1}{2} -\sin^2\theta_W\right) 
m_Z^2\cos 2 \beta\cr
\mer^2 &=& m^2 + 0.15M^2  -\sin^2\theta_W m_Z^2\cos 2\beta ,
\end{eqnarray}
and $\mlsp\sim M_1$ and $\mzii\sim M_2$ if $M\ll\mu$.  In \cite{HP},
it was shown that the two edges can be observed separately even if
$\zii\rightarrow \sel$ is not strongly phase space suppressed. It
might also be possible to find evidence for light left--handed
sleptons by looking into the relative strengths of different SUSY
signals. If $\zii\rightarrow \sel, \snu$ is open, $\wi \rightarrow
\sel$ or $\snu$ is also open and dominant, yielding relatively large
$l^+l'^-$ and $l^+l^-l'$ signals compared to the $l^+l^-$ signal.

If we assume bottom Yukawa corrections are negligible and squared
scalar masses are positive at the GUT scale, the pseudoscalar Higgs
mass $m_A$ is bounded from below as $m_A^2 > \mu^2 + m^2_{\tilde \nu}$
\cite{DN1}. Under the bino dominant assumption, and for moderate value
of $\tan\beta$, neutralino annihilation through $s-$channel poles can
thus not be important. On the other hand, for large $\tan\beta$ the
pseudoscalar Higgs boson can be light enough to achieve $\mlsp\sim
m_A/2$ \cite{DN1,DN3}. However, in mSUGRA large $\tan\beta$ also
implies a rather light $\tilde{\tau}_1$, which greatly depletes the
$l^+l^-$ signal \cite{baerbeta}; the observation of strong
multi--lepton signals would thus already indicate that $\tan\beta$ is
not very large.  Note also that for large $\tan \beta$ direct
production of the heavy neutral Higgs bosons from gluon fusion and/or
in association with $b \bar{b}$ pairs allows to detect or exclude
these Higgs bosons at the LHC for $m_A$ up to several hundred GeV
\cite{TDR}.  We will come back to the importance of determining $m_A$
later in Sec.~3.

In Fig. 1 and 2, we only looked into the parameter space with moderate
(``natural'') values of $m$ and $M$. If $m\gg M$, solutions with $\mu\sim M$
may be obtained \cite{DN3,feng}, and the assumption $\lsp
\simeq \widetilde{B}$ is not valid any more. In such a case the decay
$\zii\rightarrow \ser$ is not open. However there is still a chance
that wino--like charginos and neutralinos ($\ziv$ and $\wii$) are
produced in cascade decays, and yield additional kinematic constraints
besides the end point measurement of the $m_{ll}$ distribution from
$\zii\rightarrow\lsp ll$ decay; see Sec.~3. Large $m$ may also be
allowed when $\mlsp\sim \mz/2$, in which case $\zii \rightarrow \ser$
need not be open to make $\Omega_{\lsp}$ small.  Even then the $m_{ll}$ end
point determines $\mzii-\mlsp$ \cite{snowmass, HP1}, and the $m_{ll}$
distribution of three body $\zii$ decays is sensitive to very large
$\mer$ \cite{NY}. Another twist appears when $\zii \rightarrow \lsp Z$
is open and dominates $\zii$ decays. The small leptonic branching
ratio of the $Z$ boson might then make it difficult to study
neutralino masses, and there is no sensitivity to slepton
masses. Note, however, that in the bino--dominant region $\zii
\rightarrow Z$ decays cannot compete with $\zii \rightarrow \ser$
decays unless the latter are strongly phase space suppressed.

We already briefly alluded to the case where $\mer \gg
m_{\tilde{\tau}_1}$ due to renormalization group effects and
$\tilde{\tau}$ mixing. The lighter $\tilde{\tau}$ can be substantially
lighter than the other sleptons if $\mu\tan\beta$ is large
\cite{DN1}. In this case pair annihilation through $t$ channel
$\tilde{\tau}$ exchange can even dominate other sparticle exchange
contributions \cite{DN3}, because $\tilde{\tau}_1$ could be lighter
than the other sparticles, and the mixing induces an $S-$wave
amplitude. In \cite{HPtau} the possibility to detect and study
$\tilde{\tau}$ at the LHC is discussed. The end point of the $j_\tau
j_\tau$ invariant mass distribution, where $j_\tau$ denotes a
$\tau-$jet, is not as well determined as that of the $m_{ll}$
distribution, but it has been estimated that a $\sim 5\%$ measurement
should be possible. Even if the $j_{\tau}j_{\tau}$ end point indicates
$m_{\tilde{\tau}_1} < \mer$, the constraint on $\msq$, $\mzii$ and
$\mlsp$ from $ll$ events originating from $\zii\rightarrow \ser$
decays can perhaps be used to reduce the $m_{\tilde{\tau}}$ error, in
which case the combined error should not increase too much. However,
if $\tan\beta$ is very large, it becomes easier to have an acceptable
LSP relic density even if $\mer > \mzii$. In this case one may need a
linear collider to perform precision measurements of the nature of
$\tilde{\tau}_1$ \cite{N,NFT}, where $\sigma_{\tilde{\tau}_1}$, the
end point of the $\tau$ jet energy distribution, and a measurement of
the $\tau$ polarization would do a good job in determining the
parameters needed to predict the thermal LSP relic density.

\section{ $\Omega_{\lsp}$ in non-mSUGRA scenarios and collider signals}
 
In the previous section, we have shown that the mSUGRA assumption
predicts a bino--dominant LSP. We also found that measurements at LHC
experiments are sufficient for a prediction of $\Omega_{\lsp} h^2$, if
the cascade decay $\zii\rightarrow\ser\rightarrow\lsp$ is open and LSP
bino dominance is assumed. Now the question is if LHC experiments can
be used to check the assumption that the LSP is mostly a bino.  After
all, it is possible that $\mu$ is smaller than or of the order of the
gaugino masses, and that the mSUGRA relation $M_1 \simeq M_2/2$ is
broken. In this and the following Section, we discuss a scenario where
the inequality $M_1<M_2$ is kept, while $\mu$ is substantially smaller
than the mSUGRA prediction. In such a case $Z$ exchange effects and/or
LSP annihilation into $W$ pairs are expected to be more important than
in the strict mSUGRA scenario studied in the previous Section, and one
needs more information to predict the thermal contribution to
$\Omega_{\lsp} h^2$.

The relative size between $\mu$ and $M$ is controlled by Higgs sector
mass parameters. The MSSM Higgs potential can be written as
\begin{equation}
V = (m^2_{1}+\mu^2) H^{\dagger}_1 H_1+ (m^2_{2}+\mu^2) H^{\dagger}_2 H_2
+( B\mu H_1\cdot H_2+ h.c.) + (\rm{4th\ order\ terms}).
\end{equation} 
Here $m_1$ and $m_2$ are soft breaking Higgs masses. In the previous
Section we took $m_1=m_2=m$ at the GUT scale, which gave $\mu^2 \gg
M^2$ unless $m^2 \gg M^2$. In general, $|\mu| \sim M$ may be achieved
by allowing non-universal soft breaking Higgs masses, $m_1, \, m_2
\neq m$. For simplicity we will keep $m_1=m_2 \equiv m_h$ at the GUT
scale; we briefly comment on the effect of relaxing this assumption
below.

\begin{figure}[htbp]
\begin{center}
\includegraphics[width=6cm,angle=-90]{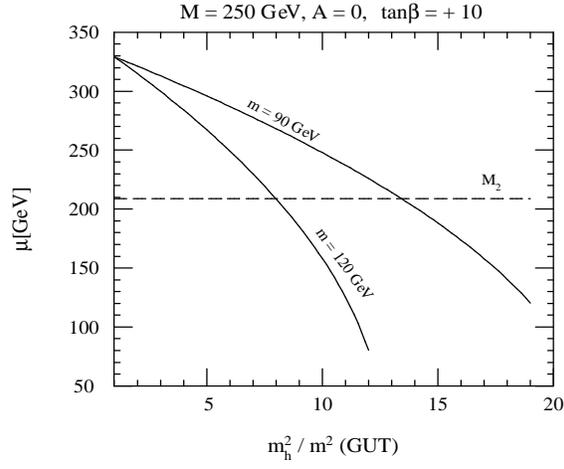}
\end{center}
\hskip 8cm a)
\begin{center}
\hskip 1cm 
\includegraphics[width=6cm,angle=-90]{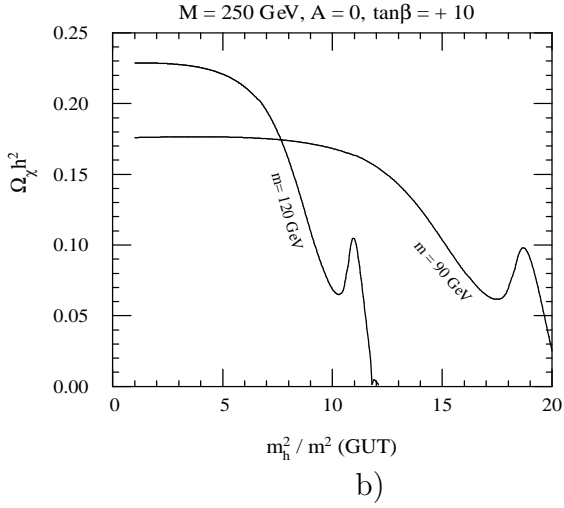}
\hskip 0cm 
\includegraphics[width=6cm,angle=-90]{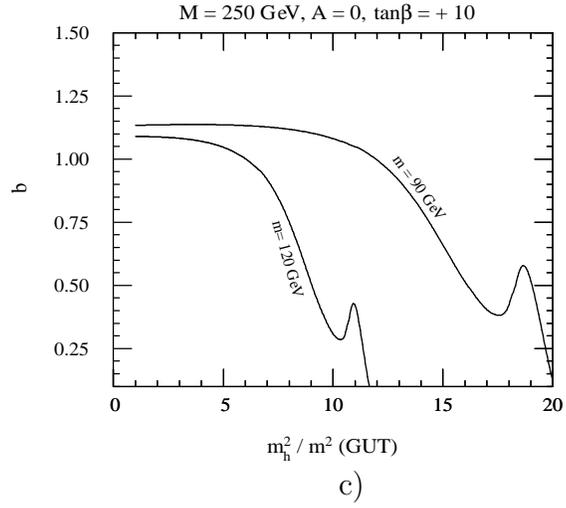}
\end{center}
\hskip 4.5cm 
b)  
\hskip 7cm 
c)
\vskip 1cm 
\caption{\footnotesize a) $\mu$ b) $\Omega_{\lsp}$ and c) $b$ 
as function of $(m_h/m)^2$. We fix $M=250$ GeV,
$A=0$ and $\tan \beta=10$.  }
\end{figure}
In Fig 3a), we plot $|\mu|$ vs. $(m_h/m)^2$.  By increasing $m_h$,
$\mu$ is reduced gradually so that $m_h^2+\mu^2$ at the GUT scale is
roughly constant. Note that the negative radiative correction to Higgs
mass $m_2^2$ is dominated by the gaugino mass through the stop and
sbottom masses for this choice of parameters\footnote{The fact that in
mSUGRA $m_2^2$ at the weak scale is almost independent of $m^2$ is
closely related to the ``focus point'' behavior studied in
ref.\cite{feng}.}.  Therefore the value of $m_h^2$ giving $|\mu| \sim
M_2$ is almost independent of $m$, $m_h\sim 330$ GeV. Generally $|\mu|
\sim M$ can be achieved if $m_h \vert_{GUT} \gsim M$; the precise
value is determined by the top Yukawa coupling. In Fig.~3b) and 3c),
we also plot $\Omega_{\lsp}$ and the $b$ factor. These quantities vary
substantially once $|\mu|$ falls below $M_2$.

\begin{figure}[htbp]
\hskip 1cm 
\includegraphics[width=6cm,angle=-90]{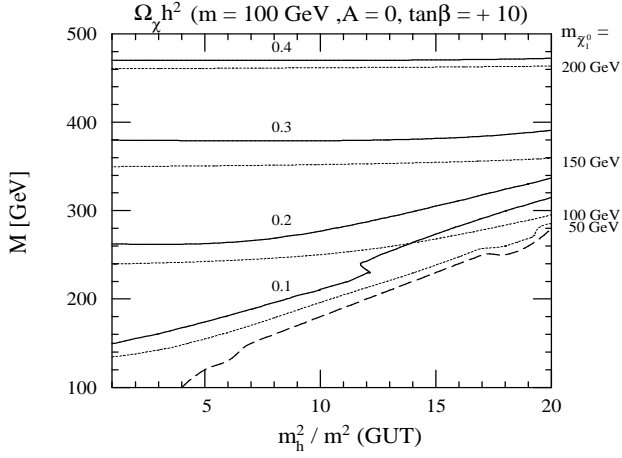}
\hskip 1cm 
\includegraphics[width=6cm,angle=-90]{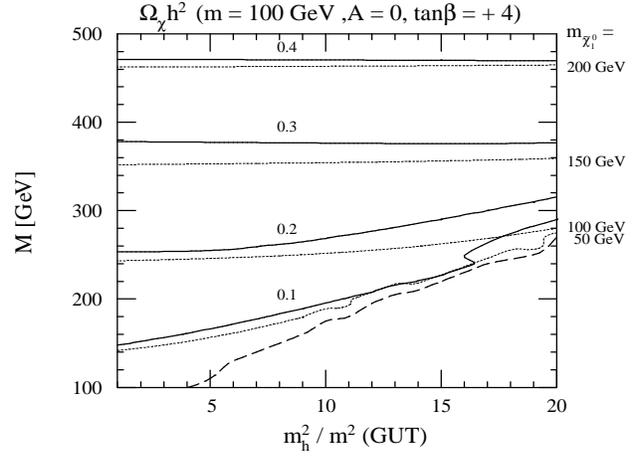}
\vskip 0.3cm 
\hskip 4.5cm 
a)  
\hskip 8cm 
b)

\hskip 1cm 
\includegraphics[width=6cm,angle=-90]{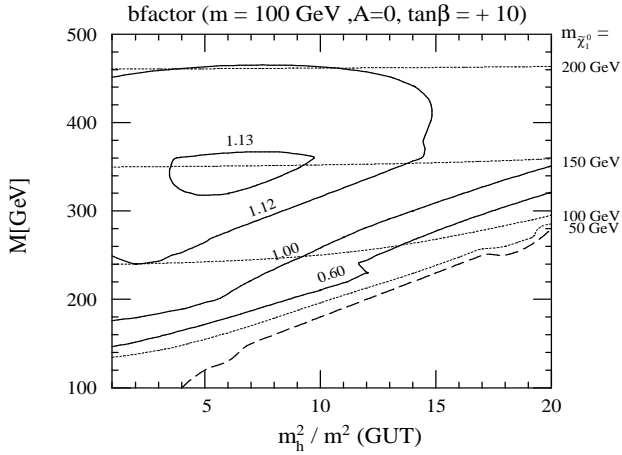}
\hskip 1cm 
\includegraphics[width=6cm,angle=-90]{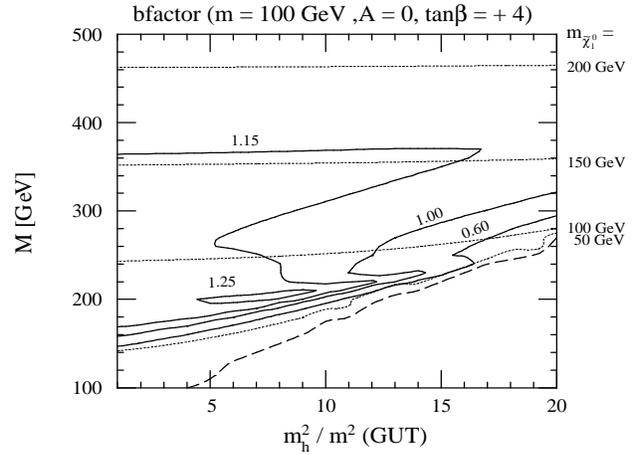}
\vskip 0.3cm 
\hskip 4.5cm 
c)  
\hskip 8cm 
d)

\caption{\footnotesize $\Omega_{\lsp}$ and the $b$ factor in the
($m_h^2/m^2, M$) plane for fixed $m$. }
\end{figure}

In Fig.~4 we show contours of constant $\Omega_{\lsp}$ and constant $b$
factor in the ($m_h^2/m^2, M$) plane for fixed $m=100$ GeV. We first
note that $\Omega_{\lsp}$ decreases as $m_h$ is increased. This is due to the
reduction of $|\mu|$ for larger $m_h$, see Fig.~3. This increases the
higgsino components of $\lsp$ and also reduces its mass. Not only
$\Omega_{\lsp}$ but also the $b$ factor decreases, therefore pair
annihilation is no longer dominated by sfermion exchanges. Especially
in Fig.~4d) we find very strong effects from LSP annihilation into $W$
pairs.  The same effect also can be found in Fig.~3b and c), where the
rise of $b$ and $\Omega_{\lsp}$ corresponds to the closure of the $WW$
mode.  Note that no consistent solution with electroweak symmetry
breaking exists below and to the right of the dashed
line.\footnote{Right on this line electroweak symmetry breaking
requires $\mu = 0$. Searches for neutralino and chargino production at
LEP therefore exclude the region just above the dashed line. However,
this experimentally excluded region is very narrow, since $|\mu|$
varies very rapidly near the maximal allowed value of $m_h$, as shown
in Fig.~3.}

As mentioned earlier we assume the two soft breaking Higgs masses to
be the same at the GUT scale. However, once $\tan^2 \beta \gg 1$, the
higgsino mass $|\mu|$ is essentially only sensitive to the value of
$m_2^2$. One can therefore increase or reduce the pseudoscalar Higgs
mass by varying $m_1^2$ at the GUT scale without affecting $\mu$ at
the weak scale significantly. Nevertheless, as long as $m_1^2(M_X) >
0$ and $\tan \beta$ is not very large, $m_{A} = \sqrt{ m_1^2 + m_2^2 +
2\mu^2 }\vert_{\rm weak}$ remains well above $2\mlsp$. However, in
principle there is nothing wrong with having $m_1^2(M_X) < 0$, as long
as the ``boundedness condition'' $m_1^2 + m_2^2 > 2 |B \mu|$ remains
satisfied at all scales. If $m_A \simeq 2 \mlsp$, the thermal LSP
relic density is very small. Strictly speaking the constraints on
$\Omega_{\lsp} h^2$ that we will derive below are therefore merely upper
bounds as long as we cannot prove experimentally that $m_A$ is well
above $2 \mlsp$.

The reduction of $|\mu|$ would alter SUSY signals at colliders
significantly. When $|\mu| \lsim M$, $\ziv$ and $\wii$ production from
the decay of $SU(2)$ doublet squarks becomes important as they have
substantial wino component. This leaves an imprint on the kinematics
of di--lepton events, which gives us access to additional MSSM
parameters, especially when the decay channels of neutralinos and
charginos into real sleptons are open. This increases the statistics
of clean $ll$ +jets+missing $P_T$ events, since the channels
\begin{eqnarray}
\wii &\rightarrow&  \snu^{(*)}\rightarrow \wi \cr
\ziv&\rightarrow  & \sel^{(*)}(\ser^{(*)}) 
\rightarrow \zii(\lsp)
\end{eqnarray}
should be seen in addition to the conventional $\zii \rightarrow \lsp
ll$ signal.\footnote{ When $M_1$ and $M_2$ have the same sign and
$|\mu|$ is not too small, one of the neutralinos is very
higgsino--like and would not be produced from the first and second
generation squark decays. Here we implicitly assume $M_1 < \mu
\lsim M_2$, in which case the higgsino--like state is $\ziii$. }

As a result, 
\begin{enumerate}
\item 
The $m_{ll}$ edge, and the other end points of the $jl$ and $jll$
invariant mass distributions of $\wii\rightarrow \snu \rightarrow \wi$
may be measured. By identifying $\tau-$leptons from $\wi$ decay, one
can confirm experimentally that the cascade decay originates from
$\wii$.  This gives lower {\em and} upper bounds on $|\mu|$, which in
turn constrain the size of the higgsino component of the LSP. An
explicit example will be analyzed in Sec.~4.
\item $\ziv$ may decay into $\sel$ directly followed by $\sel$ decay
into $\zii$ or $\lsp$. On the other hand the decay $\zii\rightarrow
\sel$ is usually forbidden or kinematically suppressed. Therefore
$\zii$ decays and $\ziv$ decays give information on different slepton
masses.
\end{enumerate}
         
Note that there are substantial constraints on --ino masses and
slepton masses from $SU(2)\times U(1)$ gauge invariance. The six
chargino and neutralino masses are determined (up to radiative
corrections \cite{pierce}) by the values of the four parameters $M_1$,
$M_2$, $\mu$ and $\tan\beta$, while $\mel$ and $\msnu$ are related by
\begin{equation} 
\msnu^2-\mel^2=m_Z^2 \cos^2\theta_W \cos2\beta
\end{equation}
Therefore the measured edges and end points originating from several
decay chains can over--constrain the relevant MSSM parameters.

\begin{figure}[htbp]
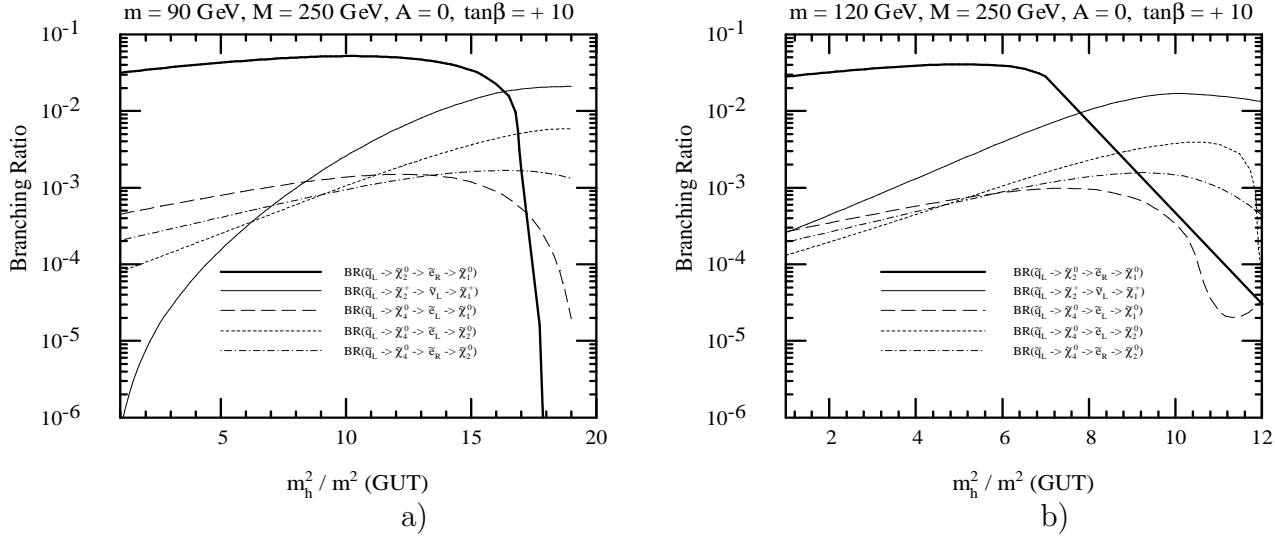

\includegraphics[width=6.5cm,angle=-90]{sig1.epsf}
\hskip 0.5cm 
\includegraphics[width=6.5cm,angle=-90]{sig2.epsf}
\vskip 0.3cm 
\hskip 3.8cm 
a)  
\hskip 8cm 
b)
\caption{\footnotesize 
Squark decay branching ratios for a) $m=90$ GeV 
and b) $m=120$ GeV, for $M=250$ GeV, $A=0$ and $\tan\beta=10$.
We average $\tilde{u}_L$ and $\tilde{d}_L$ decay branching 
ratios.}
\end{figure}

In Fig. 5, we show various $\tilde{q}_L$ decay branching ratios,
defined as an average of $\tilde{u}_L$ and $\tilde{d}_L$ branching
ratios.  As $m_h$ increases we find substantial branching ratios into
the heavier neutralino and chargino, once $|\mu|$ becomes comparable
to $M_2$. The sources of the rise of the of $\wii$ signal are:
\begin{enumerate}
\item The increase of squark branching ratios into $\wii$ and $\ziv$
with increasing wino component. In the limit $|\mu| \ll M_2$ the
branching ratios satisfy (assuming $M_1 < M_2$):
\begin{equation}
Br(\tilde{q}_L\rightarrow \wii): Br(\tilde{q}_L \rightarrow \ziv): 
Br(\tilde{q}_L \rightarrow \wi)= 2:1:0.
\end{equation}
Throughout Fig.~5, $\ziii$ is a nearly pure higgsino and does not have
a substantial branching ratio.  See Table~2 in Sec.~4 for an explicit
example.

\item The growth of the $\wii$ decay branching ratio into 
$\snu l$, again due to the increase of its wino 
component. Note that $\wii\rightarrow \snu$ is kinematically 
favored compared to $\wii\rightarrow\sel$. 

\item There is little or no phase space for $\snu$ decays into $\wi l$
if $M_2 < |\mu|$. On the other hand, this mode may open up when $\wi$
becomes higgsino--like. Indeed for Fig. 5a), in the region of small
$m_h$ the decay $\snu \rightarrow \wi$ is suppressed due to the near
mass degeneracy between $\snu$ and $\wi$.

\item 
Finally, ``conventional'' modes such as $\sq \rightarrow \wi$ and $\sq
\rightarrow \zii$ will be suppressed due to the reduced gaugino
components of these light states.  Moreover, the decay
$\zii\rightarrow \ser$ may be suppressed or closed kinematically even
for small values of $m$, if $|\mu| < M_2$. In the limit $|\mu| < M_1$,
$\mzii - \mlsp$ is small, and the corresponding $m_{ll}$ distribution
might be too soft to be accessible experimentally.

\end{enumerate}

These observations tell us that one should look for $\wii$ and $\ziv$
signals in addition to conventional $\zii \rightarrow \lsp l^+l^- $
decays. Discriminating experimentally between scenarios with $|\mu| >
M$, where these new signals are small, and $|\mu| < M_2$, where they
are expected to be significant, would be important to predict the mass
density of the Universe. In the next Section we illustrate how these
new signals could be analyzed at the LHC.

\section{ Analyzing the MSSM with $|\mu| \sim M$ at the LHC}
\subsection{A Monte Carlo Study}

\begin{table}[hbt]
\begin{center}
\begin{tabular}{|c|c||c|c||c|c||c|c|}\hline
 $m$  & 90.0  & $\tan\beta$ & 10 & $m_1$ & 360 &$m_2$ & 360 \\\hline
 $M$ & 250  &  $\mu$& 199.85 &
 $M_1$ & 103.9  &  $M_2$& 208.75\\ \hline
\hline
 $\mer$ & 139.3  & $\mel$ & 206.09  & $\mupl$ &556.07 &
 $\mdnl$ & 561.67 \\ \hline
 $\msnu$ & 190.28  &  $\mlsp$ & 93.18  &$\mupr$  
& 534.36 & $\mdnr$ & 533.21\\ \hline
 $\mzii$ & 155.13  & 
 $\mziii$ & 208.74  &$\mwi$ & 148.44 & $\mwii$ & 272.52\\ \hline
 $\mziv$ & 273.8  & $\mntau$ &188.67 &
 $\mtl$ & 374.43 & $\mth$ & 563.81\\ \hline
 $\mtaul$ & 132.56 & $\mtauh$ & 206.06 &
 $\mbl$ & 498.27 & $\mbh$ & 531.2 \\ \hline
 $m_h$ & 112.59 & $m_P$ & 436.98 &  $m_H$& 437.63& $m_{\tilde{g}}$& 624.36 
\\ \hline
\end{tabular}
\end{center}
\caption{\footnotesize Mass parameters and relevant sparticle
masses in GeV for the point studied in this paper. ISAJET \cite{ISAJET} was 
used to generate this spectrum.} 
\end{table}

We now study leptonic SUSY signals at the LHC for a case where $\wii$
production from $\tilde{q}_L$ decays is sufficiently common to be
detectable. We used ISAJET 7.42 \cite{ISAJET} to generate signal
events, while ATLFAST 2.21 \cite{ATL} was used to simulate the
detector response. For this analysis we choose the MSSM parameter
point shown in Table~1. Here we took a moderate value for $M$, leading
to a large sample of signal events.
 
The value of the GUT scale Higgs mass is chosen such that the
$\tilde{B}$ component of $\lsp$ $N_{\tilde B}=0.9$, so that effects
from its other components on the predicted LSP relic density start to
be significant; $(N_{\tilde B}, N_{\widetilde W},$ $ N_{{\tilde H}_1},
N_{{\tilde H}_2})$ $ =(0.91, -0.15,$ $
0.19, -0.35)$. In ISAJET this requires $m_h/m=4$. Our one--loop RG
analysis described in Sec.~3 reproduces this point for $m_h/m\sim
3.74$, see Fig.~3a. See also Figs.~3b) and 3c) for the corresponding
values of $\Omega_{\lsp}$ and the $b$ factor. Reducing $|\mu|$ even further
(by increasing $m_h$) would lead to even larger differences to
well--known mSUGRA scenarios, making it easier to measure all relevant
parameters.

\begin{table}[htb]
\begin{center}
\begin{tabular}{|c|c||c|c|}
\hline
$\sul\rightarrow\zii$& 20.7&    $\sdl\rightarrow\zii$ &14.5 \\ \hline 
$\sul\rightarrow\ziii$& 0.4 &  $\sdl\rightarrow\ziii$& 0.7 \\ \hline 
$\sul\rightarrow\ziv$& 12.2  & $\sdl\rightarrow\ziv$& 15.4 \\ \hline 
$\sul\rightarrow\wii$& 21.4&   $\sdl\rightarrow\wii$& 36.4 \\ \hline 
$\zii\rightarrow\ser$ & 23.6 & $\ziv\rightarrow\sel$ & 5.5 \\ \hline 
$\wii\rightarrow \snu$ & 9.7 & $\ziv\rightarrow\ser$ & 1.1 \\ \hline 
$\wi\rightarrow\staul$ & 89.1 &  $\snu\rightarrow\wi$ & 40.1\\ \hline
$\sel\rightarrow\zii$ & 38.2 & $\ser\rightarrow\lsp$ &100 \\ \hline
$\sel\rightarrow\lsp$ & 19.8 & & \\ \hline
\end{tabular}
\end{center}
\caption{\footnotesize Relevant branching ratios in \% for our sample 
parameters.}
\end{table}

In $pp$ collisions one mostly produces squarks and gluinos.  They
decay further into neutralinos and charginos.  In our case gluinos do
not decay exclusively into third generation squarks; the branching
ratio into first and second generation left handed squarks is about
20\% (11.3\% into $\tilde{u}_L$ and 9.6\% into $\tilde{d}_L$).
These squarks then often decay into $\wii$ and $\ziv$. The branching
ratios relevant for the following discussions are summarized in
Table~2. The tiny branching ratio into $\ziii$ is due to the fact
that it is mostly higgsino. In Table~3 we show the dominant cascade
decay processes which produce opposite sign same flavor lepton pairs
in the final state. In this Table we also list the corresponding end
points of the kinematic distributions discussed in Sec.~2, see
eqs.(\ref{ekin1}) and (\ref{ekin2}).

\begin{table}
\vskip 1.5cm 
\begin{center}
\begin{tabular}{|l||c|c|c|c|c|}
\hline
mode& $m^{\rm max}_{jl}$ & $m^{\rm min}_{jl}$ & $m^{\rm max}_{jll}$ & 
$m^{\rm min}_{jll}$& 
$m_{ll}^{\rm max}$  
\\ \hline
\hline
D1)\ $\sql\rightarrow\zii\rightarrow\ser\rightarrow\lsp$
& 400.3 &237.0 &430.6 &164.2 &50.6
\\ \hline
D2)\ $\sql\rightarrow\wii\rightarrow\snu\rightarrow\wi$ 
&351.0 & 260.0 &411.1 &211.1 &121.9
\\ \hline
D3)\ $\sql\rightarrow\ziv\rightarrow\sel\rightarrow\zii$
& 322.7 &269.5&403.3 &207.5 &117.9
\\ \hline
D4)\ $\sql\rightarrow\ziv\rightarrow\sel\rightarrow\lsp$
& 437.0& 321.4& 460.6& 246.6& 159.6
\\ \hline
D5)\  $\sql\rightarrow\ziv\rightarrow\ser\rightarrow\lsp$
& 421.5 & 292.2 &460.6 & 270.3 &174.3
\\ \hline
\end{tabular}
\end{center}
\caption{\footnotesize End points or edges of invariant mass
distributions (in GeV) for different decay processes.}
\end{table}

We now show several SUSY event distributions, after applying the
following cuts \cite{HP1} to reduce the SM background to a negligible
level:
\begin{itemize}
\item 4 jets with $P_{T,1}> 100$ GeV and $P_{T,2,3,4}> 50$ GeV. 
\item $M_{\rm eff} \equiv$ $ P_{T,1} +P_{T,2} +P_{T,3} +P_{T,4} + \esla_T$
$ > 400$ GeV. 
\item $\esla_T > {\rm Max}( 100 \, {\rm GeV}, 0.2 M_{\rm eff})$.
\item Two isolated leptons with $P^l_T>10$ GeV and $\vert\eta\vert<2.5$.
Isolation is defined as having less than 10 GeV energy deposited
in a cone with $\Delta R = 0.2$ around the lepton direction.
\end{itemize}

In the following plots we reduce SUSY backgrounds by subtracting event
samples with different flavor, opposite sign dileptons ($e^+\mu^-$ and
$\mu^- e^+$ ) from the sum of the $e^+e^-$ and $\mu^+\mu^-$ event
samples. To do this consistently we require two and only two isolated
leptons in the final state.\footnote{However, as we have discussed in
previous Sections, the rates of 4 and 3 lepton events compared to 2
lepton events must contain important information about MSSM
parameters.} We generated events corresponding to an integrated
luminosity of 200 $fb^{-1}$, but the figures are normalized to 100
$fb^{-1}$.

\begin{figure}[htbp] 
\begin{center}
\includegraphics[width=8cm]{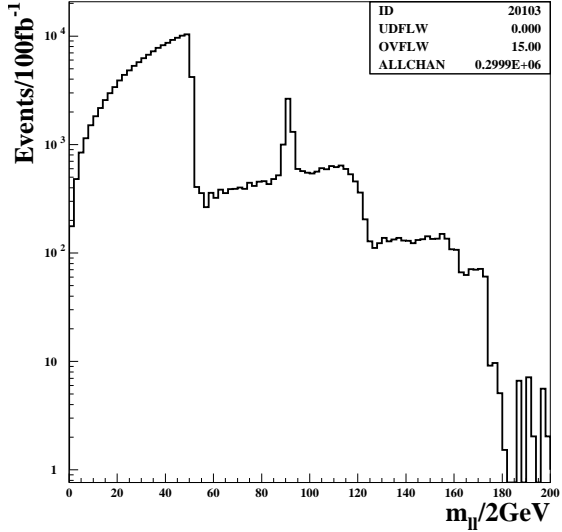}
\end{center}
\caption{\footnotesize The $m_{e^+e^-} + m_{\mu^+\mu^-} - m_{e^+\mu^-}
- m_{e^-\mu^+}$ distribution for the parameter point listed in
Table~1.}
\end{figure}

In Fig.~6 we show the di--lepton invariant mass distribution for our
representative point.  After the subtraction of $e\mu$ events, we see
a distribution with at least four edges.\footnote{Note that at
$m_{ll} \sim 55$ GeV, this subtraction reduces the number of events by
a factor of 0.35. The fluctuation of the resulting distribution is
therefore higher than what is expected from the number of events in
this distribution.} They are consistent with those found in
Table~3. Note that a rather weak edge from decay D3) should appear
very close to the one from D2) in both the $M \gg |\mu|$ and $|\mu|
\gg M$ limit; $\mwii \sim \mziv$, $\mzii \sim \mwi$ and $\msnu \sim
\mel$ hold in a wide region of parameter space. The two edges must be
separated out by fitting the smeared $m_{ll}$ distribution. Note that
since the kinematics of two decay chains is expected to be similar,
the systematic errors associated with the fitting should be small. It
seems that at least the first four $m_{ll}$ edges can be used for the
fit of MSSM parameters, while it is not clear if the last one is
detectable statistically.

We then follow the analysis of \cite{HP}, by taking the jets with the
first and the second largest $P_T$ and considering their $m_{jll}$
distributions. We label $j_1$ and $j_2$ so that $m_{j_1ll} <
m_{j_2ll}$. We then find that most events have $m_{j_1ll}$ below $\sim
400$ GeV. The $m_{jll}$ distribution will contain events from the
different decay chains listed in Table~3, but they can easily be
separated out by requiring $m_{ll}$ to lie between certain values. For
example, if we require that $m_{ll} < 55$ GeV (Fig.~7a) and $55
<m_{ll} < 125$ GeV (Fig.~7b), the distributions should dominantly
contain events from decay chains D1) and D2), respectively. In Fig.~7a)
the $m_{jll}$ end points are indeed consistent with the values of end
points listed in Table~3. The distribution in Fig.~7b) is somewhat
smeared out near the end point, due to contamination from $\ziv$
decays.

\begin{figure}[htbp]
\begin{center}
\hskip -0.5cm 
\includegraphics[width=7.5cm]{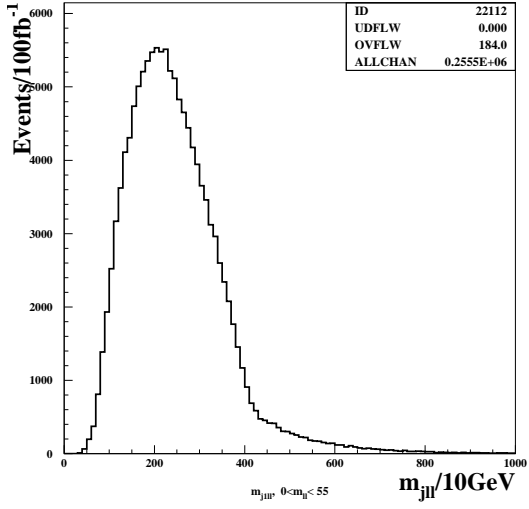}
\hskip 0cm 
\includegraphics[width=7.5cm]{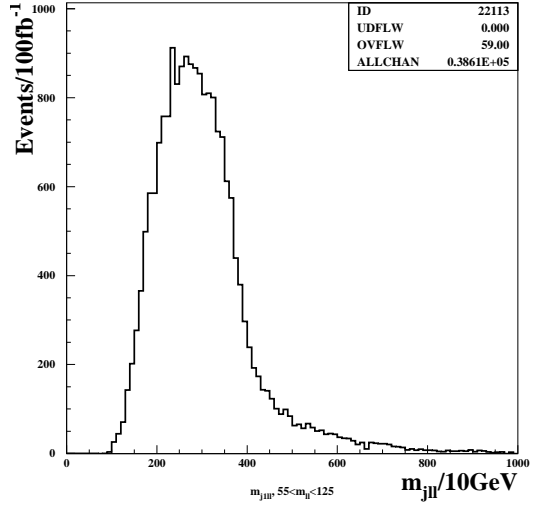}
\end{center}
\vskip-0.3cm 
\hskip 4cm 
a)  
\hskip 7.5cm 
b)
\begin{center}
\hskip -0.5cm 
\includegraphics[width=7.5cm]{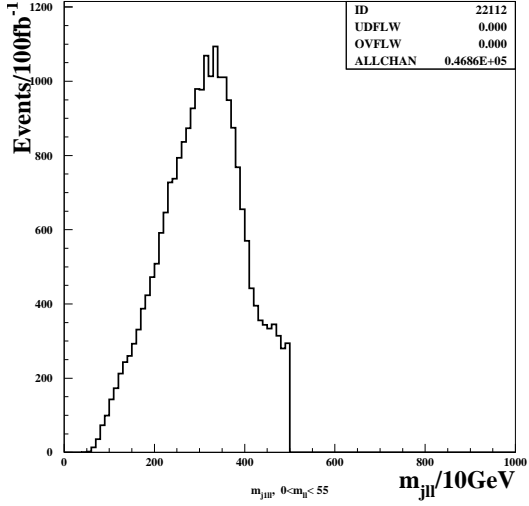}
\hskip 0cm 
\includegraphics[width=7.5cm]{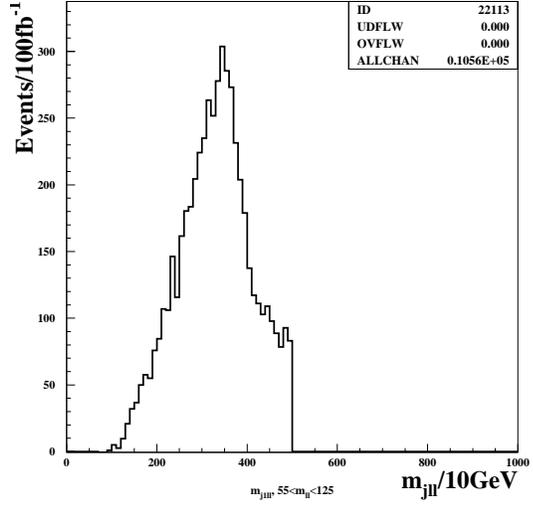}
\end{center}
\vskip-0.3cm 
\hskip 4cm 
c)  
\hskip 7.5cm 
d)

\caption{\footnotesize a), b): $m_{j_1ll}$ distributions for a)
$m_{ll}<55$ GeV, b) $55$ GeV $<m_{ll} < 125$ GeV. c), d): The same
distributions after requiring $m_{j_1ll} < 500$ GeV $<m_{j_2ll}$.}
\label{fig7}
\end{figure}

In the next step we select events where $m_{j_1ll} < 500 \, {\rm
GeV}<m_{j_2ll}$; the resulting $m_{jll}$ distributions are shown in
Figs.~7c) and 7d). These additional cuts have been applied in
ref.\cite{HP} because they reduce the probability to select the
``wrong'' jet, which does not come from primary $\tilde{q}_L$ decays.
The $m_{j_1 ll}$ distribution is then substantially harder, better
reflecting the distribution of the ``correct'' jet. Especially for
events with 55 GeV$ < m_{ll} < 125$ GeV, the $m_{j_1ll}$ end point of
decay D2) can be seen more clearly over the distributions from $\ziv$
decays D4) and D5), which have higher $m_{j_1ll}$ edges.

\begin{figure}[htbp]
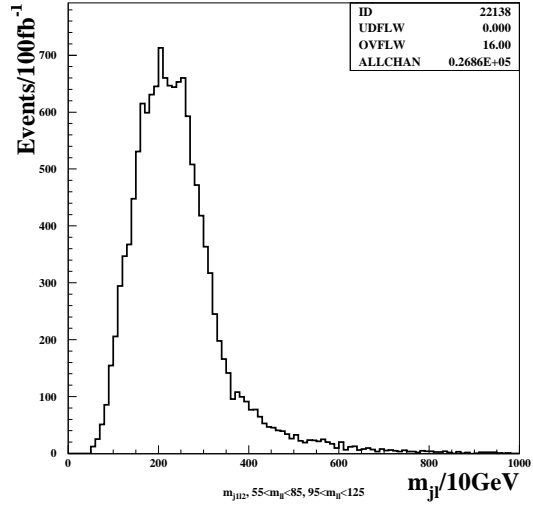

\begin{center}
\hskip -0.5cm 
\includegraphics[width=7.5cm]{mjll2_32_2.epsf}
\hskip 0cm 
\includegraphics[width=7.5cm]{mjll2_32_3_zcut.epsf}
\end{center}
\vskip-0.3cm 
\hskip 4cm 
a)  
\hskip 7.5cm 
b)
\begin{center}
\includegraphics[width=7.5cm]{mjll0_32_3_zcut.epsf}
\end{center}
\vskip-0.3cm 
\hskip 7.5cm 
c)  
\caption{\footnotesize Distribution of the higher of the two
$m_{{j_1}l}$ values for a) 0 $<m_{ll}<55$ GeV, b) 55 GeV$<m_{ll}<$85
GeV or $95$ GeV $<m_{{j_1}l}<125$ GeV with the cuts $m_{{j_1}ll}<500$
GeV$<m_{{j_2}ll}$. Figure c) corresponds to b) without the cuts on
$m_{j_1ll}$ and $m_{j_2ll}$.}
\label{fig8}
\end{figure}

In Figs.~7c) and d), we nevertheless see some continuous background
near the end point of $m_{j_1ll}$ which cannot be explained by $\ziv$
contamination. Note that for our choice of parameters, $\sql$ is
considerably lighter than for the case studied in \cite{HP}; moreover,
$\mwii$ is not too small compared to $\msq$. The probability
that one of the two hardest jets does not come from primary squark
decays should therefore be higher than in the example analyzed in
ref.\cite{HP}.

These mis-reconstructed events also contaminate the $m^{\rm
max}_{j_1l}$ edge if we demand $m_{j_1ll} < 500 \, {\rm GeV} <
m_{j_2ll}$, as can be seen in Figs.~8a), b). Here we plot the higher
of the two $m_{j_1l}$ values in each event.  However, the edges seem
to be higher than the expected values in Table~3.  Note that we
exclude events with $m_{ll} \sim m_Z$ because $\wii \rightarrow Z\wi$
followed by $ Z \rightarrow ll$ has a higher $m_{j_1l}$ edge.  For the
sample with $m_{ll}<55$ GeV, the contamination is seen as a change of
slope, while for the samples with $55$ GeV$<m_{ll}<85$ GeV or $95$
GeV$<m_{ll}<125$ GeV, no structure can be seen near the expected end
point.

This contamination actually was to be expected, because the events
that fall above the real $m_{j_1ll}$ edge must be mis-reconstructed
events where the jet originates from another sparticle decay or QCD
radiation. Therefore the corresponding $m_{jl}$ has no need to respect
$m^{\rm max}_{jl}$, either; it tends to have a value larger than this
nominal end point. The artificial upper limit of the $m_{j_1ll}$
distribution imposed by the cut then distorts the event distribution
in Fig.~8 b). We find that the $m_{j_1 l}$ distribution {\em without}
the requirement $m_{j_1ll}<500 \, {\rm GeV}<m_{j_2ll}$ reproduces the
$m^{\rm max}_{jl}$ end point of decay chain D2) better for the events
with $125$ GeV$>m_{ll}>95$ or $85$ GeV$> m_{ll} > 55$ GeV (Fig.~8c),
although the distribution is still affected somewhat by events coming
from $\ziv$ decays.

The lower edge $m_{jll}^{\rm min}$ may be reconstructed from the
$m_{{j_2}ll}$ distribution requiring $m_{ll} \geq m_{ll}^{\rm
max}/\sqrt{2}$; see Fig.~9. This distribution is much harder than the
corresponding $e\mu$ distribution; this is a sign that the observed
lower edge is real. The fit of the end point distributions will be
given elsewhere \cite{second}.

\begin{figure}[htbp]
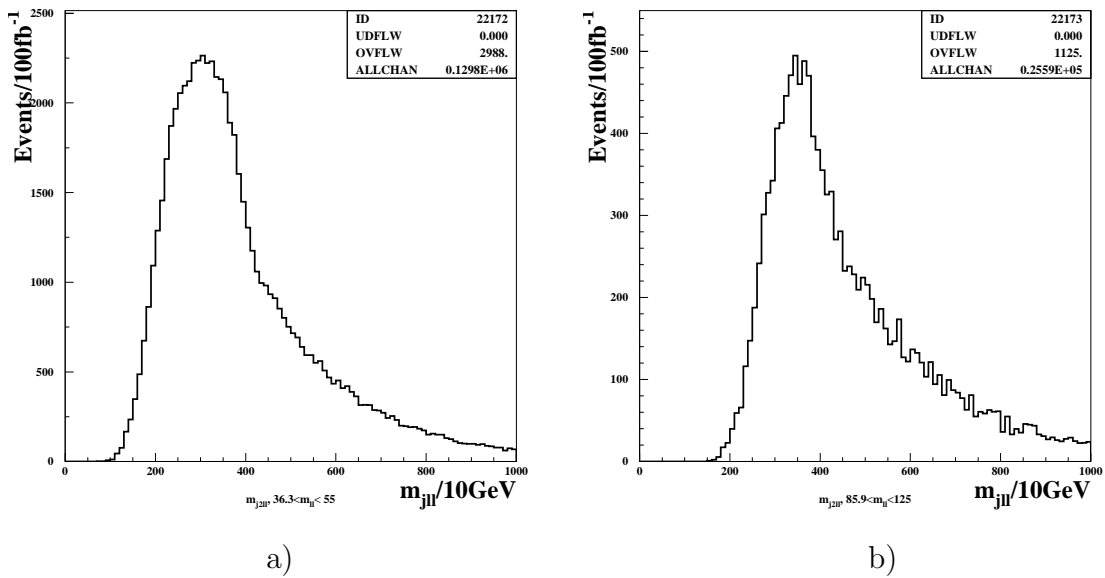

\begin{center}
\hskip -0.5cm 
\includegraphics[width=7.5cm]{mjll0_30_2.epsf}
\hskip 0cm 
\includegraphics[width=7.5cm]{mjll0_30_3.epsf}
\end{center}
\vskip-0.3cm 
\hskip 4cm 
a)  
\hskip 7.5cm 
b)

\caption{\footnotesize $m_{j_2ll}$ distribution for a) $36.3$ 
GeV$<m_{ll}<55$ GeV and b) $85.9$ GeV$<m_{ll}<125$ GeV.  }
\label{fig9}
\end{figure}

\begin{figure}[htbp]
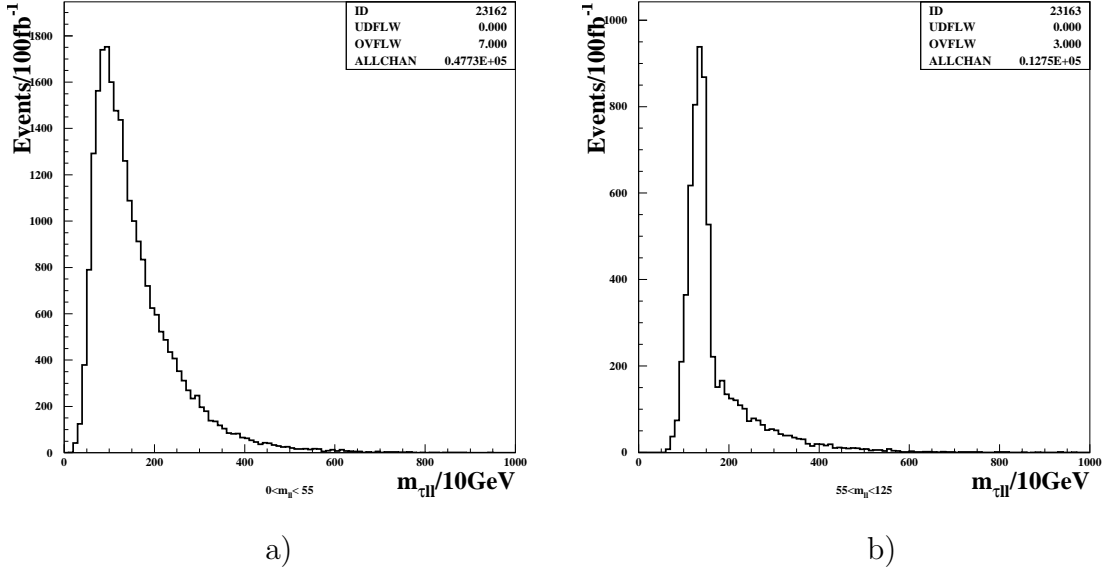

\begin{center}
\hskip -0.5cm 
\includegraphics[width=7.5cm]{mjll0_24_2.epsf}
\hskip 0cm 
\includegraphics[width=7.5cm]{mjll0_24_3.epsf}
\end{center}
\vskip-0.3cm 
\hskip 4cm 
a)  
\hskip 7.5
cm 
b)

\caption{\footnotesize The invariant mass distribution 
$m_{j_{\tau}ll}$ for a) $m_{ll}<55$ GeV and b) for 55 GeV$<m_{ll}<125$ GeV.
}
\label{fig10}
\end{figure}

We now discuss the possibility to identify $\wii$ decays through the
chain D2).  In this case most daughter $\wi$'s would decay further as
$\wi \rightarrow \staul \rightarrow \lsp$, see Table~2, producing a
$\tau$ lepton in the last step of the cascade decay.  The $\tau ll$
invariant mass never exceeds $\mwii-\mlsp$.  Hadronic $\tau$ decays
might be identified by looking for a narrow jet that is isolated from
other jet activity. Instead of studying the jet selection, we use
information from the event generator to chose jets consistent with the
parent $\tau$ direction in the event list.  The jet with minimum $dR$
is selected as $\tau-$jet if $dR<0.3$, $|\eta|<2.5$ and
$P_T/P_{Tj}>0.9$, where $P=P_{\tau}-P_{\nu_{\tau}}$.  We plot the
$m^{\rm min}_{j_{\tau}ll}$ distribution, where $j_{\tau}$ is selected
so that $m_{j_{\tau} ll}$ is minimal if the event contains several
$\tau-$jets.  When we compare the distributions for $m_{ll}<55$ GeV
(Fig.~10a) and $55$ GeV $< m_{ll} < 125$ GeV (Fig.~10b), we find the
latter events clustered in the region $m_{j_{\tau}ll}<190$ GeV, while
no such structure is found for the events with $m_{ll}<55$ GeV. The
only possible interpretation would be that most $ll$ pairs with 55
GeV$< m_{ll} <$125 GeV stem from the decay of a charged particle,
$\wii$.

In the above plot, we are assuming 100\% acceptance of $\tau$ jets and
no contamination from QCD jets. A rejection factor of $O(10^2)$
against QCD jets together with a 40\% $\tau$ identification efficiency
might be possible in the ATLAS experiment for jets with $P_T>30$ GeV
\cite{TDR}. In \cite{HPtau}, the fake tau distribution is studied
assuming a rejection factor of 15 for the case where $\zii$ decays
dominantly into $\tau^+\tau^-\lsp$. Fake $\tau$ backgrounds are then
sizable in the region above the edge of the signal $\tau^+\tau^-$
distribution. The use of the $j_{\tau} ll$ distribution to clean up
the $\wii$ sample might nevertheless help to reconstruct edges from
decay chain D2).

\subsection{Parameter Fitting}

In the previous subsection, we checked if it is possible to
reconstruct the end points of invariant mass distributions involving
charged leptons. Statistically, it seems possible to do so for decay
modes D1) and D2). This constrains mass differences among $\sql$,
$\zii$, $\lsp$, $\ser$ (D1) and $\sql$, $\wii$, $\wi$, $\snu$ (D2).
We expect that these masses can be reconstructed with O(10) GeV errors,
as was the case in ref.\cite{HP}. However, the corresponding errors on
some MSSM parameters are significantly larger.

In order to illustrate this point, we list two sets of MSSM parameters
which reproduce all kinematic end points within $\Delta\chi^2 = 1$,
where $\Delta\chi^2$ is defined as
\begin{equation}
\Delta{\chi}^2=\sum_i (M_{i}^{\rm input}-M_{i}^{\rm fit})^2/(\Delta M_{i})^2. 
\end{equation}
Here $M_i$ runs over all five end points, $m^{\rm max(min)}_{jll}$,
$m^{\rm max(min)}_{jl}$ and $m_{ll}^{\rm max}$, of the decay chains
D1) and D2) listed in Table~3.  We assume $\Delta(M_{i})$ is 1\% of
$M_{i}^{\rm input}$ for distributions involving a jet, and $0.3$ \% of
$m^{\rm max,input}_{ll}$. In Table~4, we list the solution with
maximal and minimal $\mu$ (for $\tan\beta \leq 20$) that satisfy
$\Delta\chi^2 \leq 1$.

\begin{table}
\begin{center}
\begin{tabular}{|c||c|c|c|c|c|c|c|}
\hline
&$\msql$ & $M_1$ & $M_2$ & $\mu$  & $\tan\beta$ & $\mer$ & $\msnu$ 
\\ \hline
\hline
$\mu$ max & 575.381 & 108.279 & 196.758 & 235.711 & 20.0& 147.366& 202.577
\\ \hline
$\mu$ min & 554.638 & 99.578 & 218.696 & 180.096  & 20.0 & 135.18 & 185.826 
\\ \hline
\end{tabular}
\end{center}
\caption{\footnotesize Maximal and minimal $\mu$ solution satisfying
$\Delta \chi^2\leq 1$}
\end{table}

\begin{table}
\begin{center}
\begin{tabular}{|c||c|c|c|c|c|c|}
\hline
& $\mlsp$ & $\mzii$ & $\mziii$ & $\mziv$ & $\mwi$ & $\mwii$ \\ \hline
\hline
$\mu$ max & 101.79&  163.15&  244.85& 284.94& 160.44& 285.10 \\ \hline
$\mu$ min & 89.0 & 150.97 & 190.41 & 268.72 & 144.19 & 268.74 \\ \hline 
\end{tabular}
\end{center}
\caption{\footnotesize Neutralino and chargino masses in GeV for the
maximal and minimal $\mu$ solution.}
\end{table}

Note that the errors of the dimensionful parameters are strongly
correlated, so that solutions with $\Delta \chi^2 < 1$ almost fall
onto a one--dimensional line in the seven--dimensional parameter
space. Table~4 shows that the kinematic quantities we have used in the
fit give rather weak constraints for $M_2$ and $\mu$, with errors of
order 20 GeV to 50 GeV. In fact, for fixed $\tan\beta$ we find two
distinct sets of solutions, with $\mu>M$ and $\mu<M$, respectively.
Moreover, one cannot fix the actual value of $\tan\beta$ from this
fit; one can only determine that $\tan \beta \gsim 8.65$ where the
minimum is achieved when $M_2\sim\mu$.

Table~5 shows that the corresponding chargino and neutralino masses
only vary within 15 GeV between the two extreme solutions (except for
$\ziii$, which is almost not produced in $\tilde{q}$ decays). Hence
one will need additional information, beyond the kinematics of the
decay chains D1) and D2), to reduce the errors on the fundamental
parameters.

Reducing the errors on $\mu$ and $\tan \beta$ would be necessary to
predict the thermal relic density accurately. The $\mu$ (max,min)
solutions predict $\Omega_{\lsp} h^2 = 0.160$ and 0.122, respectively,
as compared to 0.152 for the input point. The $\mu$ max point predicts
a similar relic density as the input point; indeed, within the region
with $\Delta \chi^2 \leq 1$, we were not able to find solutions with
$\Omega_{\lsp} h^2 > 0.165$. On the other hand, the $\mu$ min solution
predicts a significantly smaller relic density, and even smaller
values of $\Omega_{\lsp} h^2$ are possible if we relax the upper bound
on $\tan \beta$, which was imposed ``by hand'' in this fit. For
example, there is a solution with $\tan \beta = 36$ and $(M_1, M_2,
\mu, \msq, \mer, \msnu) = (108.3, 194.7, 239.3, 576.9, 148.0, 203.9)$
(all masses in GeV), giving $\Omega_{\lsp} h^2 = 0.112$.  We hence need
to reduce the errors on {\em both} $\mu$ and $\tan \beta$. The former
determines the size of the higgsino components of the LSP, which
begins to be significant in this region of parameter space. The
product $\mu \tan \beta$ determines the amount of $\tilde{\tau}_L -
\tilde{\tau}_R$ mixing, which reduces the predicted relic density
through a reduced $\tilde{\tau}_1$ mass and enhanced $S-$wave
annihilation. In the following we discuss strategies that might be
useful for reducing the errors on these two quantities.

\begin{table}
\begin{center}
\begin{tabular}{|c||c|c|c|c|c|} 
\hline
                      &     $ \lsp$ & $\zii$&  $\ziv$ & $\wi$ &  $\wii $
\\ \hline  
\hline
$\mu$ max $\sul\rightarrow$  & $10^{-3}$ & 0.257 &0.073 & 0.553& 0.113
\\ \hline
 
     $ \sdl\rightarrow$   &   0.036 &    0.214 & 0.094&  0.412& 0.238
\\ \hline
$\mu$ min $\sul\rightarrow$ & 0.4 $10^{-3}$& 0.167& 0.162& 0.381& 0.285
\\ \hline
         $ \sdl\rightarrow$ &  0.042 &0.106 & 0.196& 0.197 &0.451
\\ \hline
\end{tabular}
\end{center}
\caption{\footnotesize Squark branching ratios for the maximal and
minimal $\mu$ solutions.}
\end{table}

One possibility is to measure some branching ratios. In Table~6, we
compare the $\tilde{q}$ decay branching ratios into charginos and
neutralinos for the two solutions. Note that the ratio of the $\wii$
and $\zii$ modes increases by more than a factor of three, from 0.45
to 1.71 for $\sul$ decay and from 1.11 to 4.25 for $\sdl$ decay, when
switching from the $\mu$ max solution to the $\mu$ min solution. This
is almost entirely due to the change of $\mu$; the value of $\tan
\beta$ is not important here (as long as $\tan^2 \beta \gg 1$). 

The relative strengths of the signals from decay chains D1) and D2)
should thus yield important information to reduce the errors on MSSM
parameters. The strengths of these signals can be extracted purely
kinematically, e.g. from the relative number of events with $m_{ll}$
below the $\zii$ and $\wii$ edge, respectively, and/or by trying to
determine the fraction of di--lepton events that have a $\tau-$jet
near the charged lepton pair, as discussed above. For a given
solution in Table~4, all chargino and neutralino mixing angles and
masses are fixed. As stated above, this is a fairly constrained fit
where all relevant sparticle masses are effectively described by one
parameter. The acceptances should then be very well calibrated from
the mass constraints, so that systematic errors should be small.

\begin{table}
\begin{center}
\begin{tabular}{|c||c|c|c|}
\hline
                       &     $\mu$ max & $\mu$ input(univ) &$\mu$ min
\\ \hline
\hline
$\zii\rightarrow\ser$  &    2$\times$ 0.06 &   2$\times$ 0.148  & 2
                       $\times$   0.105
\\ \hline
$\wii\rightarrow\snue$     &   0.063  &   0.096 &    0.109
\\ \hline
$\snue\rightarrow\wi$ &         0.433 &     0.402 &    0.380
\\\hline
\end{tabular}
\end{center}
\caption{\footnotesize Branching ratios for the maximal and minimal
$\mu$ solution. We assume universal soft sfermion masses and
$A_{\tau} = 0$  at the weak scale.}
\end{table}

In order to extract squark branching ratios from the number of events
with a lepton pair in the final state, one must know $Br(\zii
\rightarrow \ser)$ and $Br(\wii \rightarrow \snu)$. These branching
ratios also depend on MSSM parameters, as shown in Table~7.  Here we
assume that the $\stau$ and $\snut$ soft breaking mass parameters are
the same as for first and second generation sleptons.  We compute the
$\stau$ mixing angle by setting $A_\tau = 0$ at the weak scale;
whenever it is sizable, $\tilde{\tau}_L - \tilde{\tau}_R$ mixing is
anyway dominated by the contribution $\propto \mu \tan \beta$.  With
these assumptions all parameters required to compute these branching
ratios can in principle be extracted from the kinematic fitting
described above.

The least critical quantity in Table~7 is the branching ratio for
$\snue \rightarrow \wi$ decays. It decreases slightly with decreasing
$\mu$, due to the shrinking wino component of $\wi$. However, this
effect is weaker than the simultaneous increase of $Br(\wii
\rightarrow \snue)$, which is due to the increasing wino component of
$\wii$. The strength of the signal from decay chain D2) is
proportional to the product of these two branching ratios, which
varies between 0.027 and 0.041. Together with the simultaneous change
of $Br(\tilde{q}_L \rightarrow \wii)$ shown in Table~6, this means
that for our choice of parameters the signal strength of D2) decreases
rapidly with increasing $\mu$. Moreover, the relevant branching ratios
do not depend significantly on the details of the $\tilde{\tau}$
sector, and can thus be predicted fairly reliably from the quantities
listed in Table~4.\footnote{In principle $\wii$ branching ratios could
change somewhat if $m_{h_l} < \mwii - \mwi$, in which case additional
2--body decay modes of $\wii$ would open up. However, such scenarios are
already tightly constrained by LEP data, and would give rise to a
variety of Higgs signals at the LHC.}

Unfortunately this is not true for the branching ratio for $\zii
\rightarrow \ser e$ decays, which does depend strongly on the mass and
mixing angle of $\staul$. Note that the prediction in Table~7 for the
input point (0.296) differs from the input value in Table~2
(0.236). This is because we ignored the reduction of soft breaking
$\tilde{\tau}$ masses through RG effects when computing the entries of
Table~7. In the given case these effects only reduce
$m_{\tilde{\tau}_1}$ by $\sim 5$ GeV. This is sufficient to increase
the partial width for $\zii \rightarrow \staul \tau$ significantly;
note that $\zii \rightarrow \tilde{l} l$ decays are pure P--wave in
the limit $m_l \rightarrow 0$, and the available phase space for these
decays is not large in our case. Since for both fit solutions shown in
Table~4 $\tan \beta$ is significantly larger than the input value,
these solutions predict even lighter $\tilde{\tau}_1$ states and
enhanced $\stau_L - \stau_R$ mixing. The use of these parameters would
therefore underestimate the true branching ratio for $\zii \rightarrow
\ser e$ decays significantly.

In order to extract the squark branching ratios of Table~6 to better
than a factor of 2 one will therefore need additional information on
the $\stau$ sector. This might be obtained by studying $\zii
\rightarrow \tau^+ \tau^- \lsp$ decays. As mentioned at the end of
Sec.~3, it should be possible to determine the edge of the
di$-\tau-$jet invariant mass distribution to $\sim 5\%$. This would be
sufficient to detect large differences between $\stau$ and $\tilde{e}$
masses; however, it would not suffice to distinguish between the three
cases used in Table~7. To this end one would need to determine the
ratio of branching ratios for $\zii \rightarrow e^+ e^- \lsp$ and $\zii
\rightarrow \tau^+ \tau^- \lsp$ decays. The precision of this
measurement might be limited by systematic effects, since the two
signals have very different efficiencies.\footnote{Since the $\zii -
\staul$ mass difference is quite small, one may have to allow one of
the $\tau-$jets to be quite soft.} However, given that this ratio of
branching ratios differs by more than a factor of 4.5 between the
three scenarios of Table~7 we think it likely that its measurement
will help to reduce the errors of the extracted squark branching
ratios significantly. Finally, once a linear collider of sufficient
energy becomes available precision studies of $\staul$ production and
decay will be possible \cite{N,NFT}.

Let us summarize this somewhat complicated discussion by turning the
argument around. One should first extract information about the
$\stau$ sector, e.g. by comparing signals from $\zii \rightarrow
\tau^+ \tau^- \lsp$ to those from $\zii \rightarrow e^+ e^- \lsp$. This
will give information on the soft breaking masses in the $\stau$
sector as well as on the product $\mu \tan \beta$. This information,
together with the result of the kinematic fit described above, will
allow one to predict the branching ratios of the decays listed in
Table~7 with reasonable precision. This in turn will allow to
translate the measured strengths of the signals from decay chains D1)
and D2) into squark branching ratios. Finally, these branching ratios
can be used to greatly reduce the error on $\mu$.

Another way to further constrain the relevant MSSM parameters is to
include $\ziv$ decay edges from decays D3) and D4) in the fit. Just
measuring $m^{\rm edge}_{ll}$ values of these decay modes with 1\%
errors allows to reduce $\mu^{\rm max}$ to 214 GeV. However, this
would still not allow us to give an upper bound on $\tan
\beta$.\footnote{Including this new information gives the strong upper
bound $\tan \beta \leq 11$ at the $1 \sigma$ level, but $\tan \beta =
20$ remains allowed at $2 \sigma$.} Moreover, this edge may not be
visible for larger gluino and squark masses, where the production cross
section is substantially smaller.

Given that the very weak upper bound $\tan \beta \leq 20$ which we
imposed in the fit summarized in Table~4 is sufficient to predict
$\Omega^{th}_{\lsp} h^2 \simeq 0.135 \pm 0.03$, it seems certain that
the strategy outlined above will again allow to predict the thermal
relic density to better than 20\%. The only loophole occurs if $\tan
\beta$ is very large. In this case the $\lsp-\staul$ mass difference
becomes so small that the $\tau$ from $\staul$ decays become
effectively invisible at hadron colliders. At the same time $\staul -
\lsp$ co--annihilation reduces the predicted LSP relic density by up
to a factor of ten \cite{olive}. One would then need to increase $m$
and/or $M$ in order to get a cosmologically interesting value of
$\Omega_{\lsp} h^2$; $\zii \rightarrow \ser e$ decays may not be
open. In this case straightforward kinematic fitting as we described
here will not be possible at the LHC, although one should still get a
hint for the relative ordering of $\staul$ and $\ser$ masses by
observing $\tau-$jets in missing $E_T + $jets events, which will yield
the most robust SUSY signal in this case. In such a somewhat contrived
scenario kinematical precision measurements would probably only be
possible at a lepton collider.

\section{Discussion}

In this paper, we argued that LHC experiments can play a substantial
role in predicting the contribution $\Omega^{th}_{\lsp}$ of thermal
relic LSPs to the mass density of the Universe. Previous simulations
in the literature were mostly done using mSUGRA assumptions, where
usually $|\mu|^2\gg M^2$.  In such a case the measured $\zii$ cascade
decay determines $\mer$, $\mzii$ and $\mlsp$. Once we know that the
LSP is mostly bino, these three masses are sufficient to determine
$\Omega^{th}_{\lsp}$ within 20\%.

On the other hand, if Nature does not respect universality of all
scalar soft breaking masses, it is possible that $\mu\sim M$. In such
a case, one needs to know $\mu$ and $M$ in addition to $\mlsp$ and
$\mer$ to determine $\Omega^{th}_{\lsp}$, because $s$-channel exchange
of $Z$ and $h$, or $WW$ production might play important roles in
$\lsp$ pair annihilation in the early Universe.  We showed that, if
$\mu\sim M$, $\wii$ and $\ziv$ will be produced copiously in
$\tilde{q}_L$ decays. One can then determine all MSSM parameters
needed to predict $\Omega^{th}_{\lsp}$ through the study of these
cascade decay channels, using fits of kinematic end points and edges
of invariant mass distributions. The isolation of $\wii \rightarrow
\snu \rightarrow \wi$ decays by observing the subsequent $\wi
\rightarrow \tilde{\tau}$ decay may be used to improve the
reconstruction of the $\wii$ production and decay
kinematics. Moreover, the measurement of the relative number of events
from different decay chains further constrains MSSM parameters. In the
end it should again be possible to predict the thermal LSP relic
density with an error of 20\% or better even in this more complicated
scenario. In fact, this scenario is advantageous, since it allows us
to determine both the gaugino and higgsino components of the LSP;
these are needed to predict the strength of the LSP couplings to Higgs
bosons, which in turn are required for predicting the LSP--nucleon
scattering cross section. In mSUGRA scenarios with $|\mu| > M$ one can
probably only establish an upper bound on the higgsino component of
the LSP, which only allows one to derive upper bounds on LSP--Higgs
couplings.

Notice that we {\em only} used information that can be extracted from
studies at the LHC to arrive at this rather optimistic conclusion. If
any one of the relevant masses could be determined with better
precision elsewhere, e.g. at a lepton collider, the allowed region
would shrink significantly, since the fit of hypothetical LHC data
resulted in an almost one--dimensional $\Delta \chi^2 \leq 1$ domain.

In this paper we discussed the case where $m\ll M$ so that 
neutralino decay into sfermion is open. This is a good assumption 
if $\lsp$ is gaugino like, as $\lsp$ density overclose 
the Universe if $m\gg M$. For increased higgsino component of $\lsp$, 
such a requirement is no longer necessary. It is interesting to 
see if one can extract sparticle masses from decay distributions
for such cases.

In this paper, we did not study the case where $\mu \ll M_1, M_2$,
where the higgsino--like states $\lsp$, $\zii$ and $\wi$ are nearly
degenerate in mass.  In such a case we should observe $\wii, \ziv$
production from squark decays in addition to $\ziii$ production, which
is now mostly $\tilde{B}$. If scalar masses are not too large so that
decays of the heavier neutralinos and charginos into real sfermions
are open, the analysis is similar to the one that has been given in
Sec.~4. If the sfermion decay mode is closed, the decay to a
(virtual) Higgs boson might play an important role, unlike the case where
$\mu\sim M$. While on--shell Higgs bosons produced in SUSY cascade
decays can be identified \cite{TDR}, the kinematical fitting would be
more difficult since it would be entirely based on jets. Note,
however, that the thermal relic density of higgsino--like LSPs is
small unless $\mlsp \gsim 500$ GeV, in which case it might be
difficult to even discover supersymmetry at the LHC.

We also did not discuss the case where $M_2\ll M_1,\mu$ suggested in
models with anomaly mediated supersymmetry breaking \cite{RM}. These
models predict a rather heavy gluino. This results in a limited number
of events even at the LHC, making precision studies rather
difficult. In models with a not too heavy gluino while keeping $M_2\ll
M_1,\mu$, the relative number of events from $\tilde{B} \rightarrow
\ser$ and $\tilde{B} \rightarrow \sel$ might be useful to show that
$M_1\gg M_2$, if both modes are open. This would be sufficient to show
that the thermal LSP relic density is small, independent of the
relative ordering of $M_2$ and $\mu$, since both wino--like and
higgsino--like LSPs with mass in the (few) hundred GeV range
annihilate efficiently.

We thus conclude that whenever LHC experiments find a large sample of
SUSY events, it will be possible to either predict the thermal relic
density of LSPs with a fairly small error, or else one will be able to
conclude that thermal relic LSPs do not contribute significantly to
the overall mass density of the Universe. In the latter case one would
need physics beyond the MSSM, and/or a non--thermal LSP production
mechanism, to explain the Dark Matter in the Universe.

\subsection*{Acknowledgements}
The work of MD was supported in part by the ``Sonderforschungsbereich
375--95 f\"ur Astro--Teilchenphysik'' der Deutschen
Forschungsgemeinschaft.

\end{document}